# Boundary conditions for and ferromagnetic resonance spectra of magnetic bilayers coupled by interlayer Dzyaloshinskii-Moriya interactions

Elena Y. Vedmedenko[1] and Mikhail Kostylev[2]

[1] *Department of Physics, the University of Hamburg, Jungiusstr. 11, Hamburg, Germany*
[2] *Department of Physics M013, the University of Western Australia, Crawley 6009, Australia*

Interfacial Dzyaloshinskii-Moriya interaction (IF-DMI) leads to non-collinear spin configurations within the magnetic layers of multilayer heterostructures, while its interlayer counterpart (IL-DMI) minimizes chiral states between the layers. Here, we demonstrate that the symmetries of these interactions are very different, even though both arise from pairwise exchange interactions between magnetic sites mediated by nonmagnetic atoms. By deriving new boundary conditions for the exchange operator and solving the associated boundary value problem, we show that, unlike IF-DMI, which does not contribute to the FMR frequencies, IL-DMI alters the frequencies of the fundamental FMR modes and can be separated from other contributions in an FMR experiment.

I.     Introduction

Current research into condensed matter physics and magnetism is largely focused on the search for novel materials able to efficiently convert spin currents into into charge currents and vice versa [1]. In this regard, materials capable of forming non-collinear magnetic structures with unique chirality are of great interest for cutting-edge fields of Three-dimensional Spintronics and Quantum Engineering. A straightforward way to achieve these topological textures is by exploiting the presence of an antisymmetric exchange interaction known as the Dzyaloshinskii-Moriya interaction (DMI). Three main types of DMI are known: bulk DMI [2], the interfacial DMI (IF-DMI) [3], and the long-range interlayer DMI (IL-DMI) [4], which is sometimes also referred to as RKKY-DMI. The bulk DMI is responsible for forming chiral magnetic structures within the bulk of magnetic crystals [5]. The IF-DMI is induced at the interfaces of individual ferromagnetic (FM) layers with a non-magnetic metal (NM). It stabilizes non-collinear spin configurations within the magnetic layer due the breaking of symmetry caused by the presence of the interfaces [6]. The IL-DMI provides long-distance chiral coupling between magnetic layers separated by non-magnetic spacers [7-12]. All DMI subclasses are characterized by an energy term that includes a vectorial interaction strength known as the Dzyaloshinskii-Moriya vector ***D***. The vector magnitude and orientation differ among them. While the DMI in bulk materials and the IF- DMI have already been extensively investigated, the recently discovered IL-DMI has been studied in far less detail. Particularly, while it is known that the strength of IL-DMI can be comparable to that of the IF-DMI (0.1 to 0.2 mJ/m$^2$ [9,10]), tt remains an open question whether there is a fundamental difference in the symmetry breaks associated with various magnetic effects in magnetic multilayers (MML) due to the presence of IF-DMI and IL-DMI. This question is important, because both interactions arise from similar pairwise exchange interactions between magnetic sites mediated by nonmagnetic atoms.

Typically, systems exhibiting IL-DMI are identified using magnetometry and magneto-transport techniques, as IL-DMI breaks the symmetry of magnetization reversal [7-12]. However, within the paradigm of standard magnetometry measurements, separating IL-DMI from the Heisenberg-type interlayer exchange interaction (IL-HEI), which includes bi-linear (IL-LEI) and bi-quadratic exchange (IL-BEI) contributions [13-16], remains challenging. From the experiments, one typically extracts an effective value of IL-DMI, which also includes contributions from all other interlayer exchange couplings [9,10]. This limits the predictive power of the results.

One magnetic effect where the difference in symmetries of IF-DMI and IL-DMI is likely to produce readily distinguishable manifestations is Ferromagnetic Resonance (FMR) [16,17].

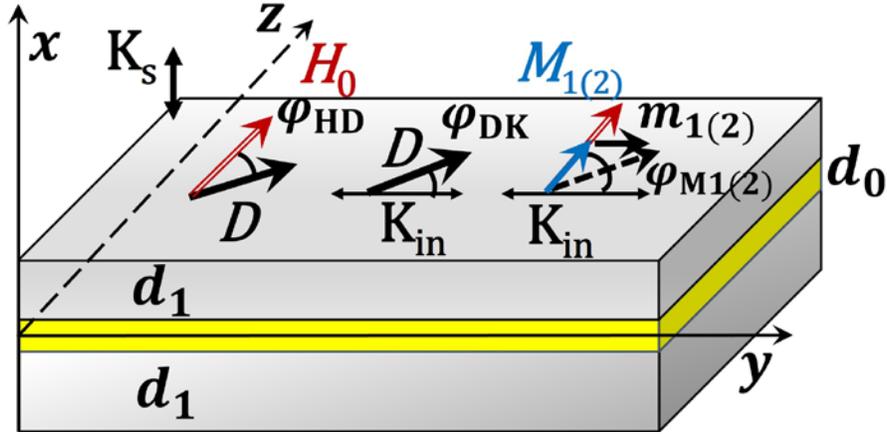

**Fig. 1** Geometry of a system consisting of two FM layers of thickness $d_1$ with magnetizations $M_i$ separated by a NM spacer of thickness $d_0$. The external magnetic field is applied along the z-direction. We consider out-of-plane anisotropy $K_s$ acting at the external surfaces and internal interfaces as well as dynamical bulk uniaxial in-plane anisotropy $K_{in}$ with angle $\varphi_{M1(2)}$ between its axis and the static magnetization vector. The IL-DMI vector $D$ and the field $H_0$ make angles $\varphi_{HD}$ and $\varphi_{DK}$ with respect to the $K_{in}$.

Additionally, FMR may provide a promising approach to separate IL-DMI from IL-HEI, as the eigenfrequency of FMR is highly sensitive to conditions at the magnetic interfaces of multilayer systems [18]. On the other hand, IF-DMI has been shown not to contribute to the FMR response due to the symmetry of this interaction [19-21].

More generally, the precessional dynamics of systems coupled by IL-DMI remain largely unexplored. The only report we are aware of focuses on the study of magnon-magnon coupling induced by IL-DMI in specific synthetic antiferromagnets [22]. This work centers on the relative orientation of magnetic layers with antiferromagnetic (AFM) IL-HEI, while the contributions of ferromagnetic (FM) and biquadratic interactions, as well as the role of interaction strength in shaping the FMR response of magnetic multilayers, were not addressed. Furthermore, the cited publication emphasizes the quantum engineering aspects of the effect, particularly concerning ultrastrong magnon-magnon coupling. However, it does not investigate the implications for material science, including the potential for extracting magnetic parameters of the layered system from FMR data.

To fill this knowledge gap, we derive interlayer boundary conditions (IL-BC) for linear magnetization dynamics in the presence of IL-DMI. These boundary conditions reveal that, unlike

IF-DMI, the IL-DMI interaction does affect the FMR response of magnetic multilayers (MML). This distinction stems from the differing symmetries of these interactions.

We observe that there is no perfect co- or anti-alignment of the dynamic magnetization vectors of the two FM layers. Such alignment would be expected if IL-HEI were the sole coupling mechanism. However, the introduction of IL-DMI disrupts this alignment. In the framework of coupled oscillations, IL-DMI introduces a phase difference between the components of the coupled system that is not an integer multiple of π. This phase shift modifies the frequencies of the fundamental FMR modes: the acoustic mode (AM) and the optical mode (OM). We then solve the boundary-value problem for the FMR dynamics of a layered structure shown in Fig. 1 using the derived IL-BC. We assume that the non-magnetic metallic spacer exhibits strong spin-orbit coupling and *sd* electronic energy contributions, leading to significant interlayer Dzyaloshinskii-Moriya interaction (IL-DMI, $D$) and interlayer Heisenberg exchange interactions (IL-HEI, $A_{12}$) between the ferromagnetic layers. We focus on the ferromagnetic (FM) configuration of the system's magnetic ground state ($A_{12} > 0$) across a broad range of IL-DMI strengths. As IF-DMI does not contribute to the FMR response, we neglect this interaction in our theoretical analysis.

In the general case, the solution of the boundary-value problem does not yield a simple explicit expression for the FMR frequencies or fields. However, by neglecting the effective fields of magnetic anisotropies and IL-HEI, we can derive a straightforward formula. For anisotropic systems with strong IL-HEI, we instead obtain the FMR frequencies or fields numerically. Additionally, we validate our findings through microscopic atomistic calculations.

We observe that there is no perfect co- or anti-alignment of the dynamic magnetization vectors of the two FM layers. Such alignment would be expected if IL-HEI were the sole layer coupling mechanism. However, the introduction of IL-DMI disrupts this alignment. In the framework of the general theory of coupled oscillations, IL-DMI introduces a phase difference between the components of the coupled system that deviates from the usual integer multiples of π. This unconventional phase shift influences the frequencies of the system's FMR modes.

We find splitting of the parent mode of the uniform magnetization precession in a single ferromagnetic layer into a family of an acoustic and optical mode for the layered structure. This happens even in the unrealistic case when IL-HEI is absent, and IL-DMI is the sole layer exchange-coupling mechanism present in the system. Additionally, we observe a very unusual angular dependence of the FMR fields: the resonance fields for the optical and acoustic modes vary in antiphase as a function of the applied-field angle relative to the vector $D$. This antiphase behavior serves as a signature of the presence of IL-DMI in the system. Ultimately, we provide suggestions for experimentalists on how to demonstrate the presence of IL-DMI in a magnetic multilayer structure using FMR.

## II Theoretical model

We proceed by solving the Landau-Lifshitz Equation of motion for the magnetization vector (LLE). The effective dynamic magnetic field in the equation includes a contribution from the inhomogeneous exchange interaction within the magnetic layers. The presence of the exchange operator necessitates proper exchange boundary conditions (EBC) at the interfaces of the magnetic layers. Therefore, an important step in our analysis is to complement the well-established Grünberg-Rado boundary conditions [23-26] with an IL-DMI term, which we derive in this work. Additionally, we incorporate volume and interfacial magnetic anisotropies into the model. The former is added as a contribution to the bulk effective field, while the latter is included in the EBC.

To make our work useful to experimentalists, we consider a configuration suitable for taking FMR absorption traces. That is, two FM layers of equal thickness $d_1$ and saturation magnetizations $M_{s1}$ and $M_{s2}$ are separated by a NM spacer of thickness $d_0$. The static magnetization vectors $\boldsymbol{M}_i$ of the FM layers lie in the film plane and are orientated along the $z$-axis, while the $x$-direction corresponds to the plane normal. We assume that $\boldsymbol{M}_i$ is forced to align along $z$ by a static magnetic field $\boldsymbol{H_0} = H_0\hat{z}$, and that $\boldsymbol{H_0}$ is strong enough to orient $\mathbf{M}_i$ precisely along the $z$-axis. The linearised LLE then takes the form as follows

$$\frac{\partial}{\partial t}\boldsymbol{m}_i = \gamma_i\,\mu_0\big[\boldsymbol{m}_i \times \boldsymbol{H}_0 + \boldsymbol{M}_i \times \boldsymbol{H}_{\text{eff},i}\big], \tag{1}$$

where $\mathbf{m}_i$ is the dynamic component of the magnetization for the layer $i$ ($|\mathbf{m}_i|<<|\boldsymbol{M}_i|=M_{si}$), $\gamma_i$ is the gyromagnetic ratio for the layer, and $\mu_0$ is permeability of vacuum. In contrast to the theory of the impact of spin pumping on the FMR response [27], where the inclusion of Gilbert damping is essential, we can safely neglect it in our case. While Gilbert damping is known to contribute to the

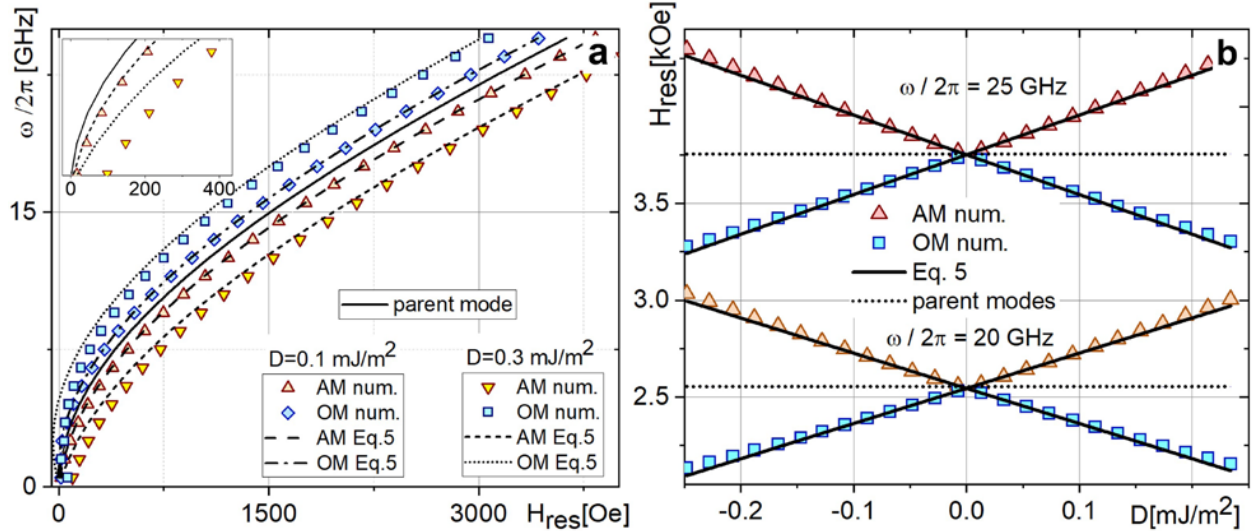

**Fig. 2** Analytical (Eq. 5) and numerical FMR spectra for two magnetic layers separated by a non-magnetic spacer and coupled by IL-DMI, with both IL-HEI and magnetic anisotropies vanishing. The analytical results are represented by lines, while the numerical results are shown by symbols. For comparison, the parent uniform mode of an uncoupled single layer is included (solid line in Panel (a) and dotted line in Panel (b)). (a) Resonance frequency $\omega/2\pi$ vs. resonance field H$_{\text{res}}$ for two $D$ values: 0.1 and 0.3 mJ/m². (b) FMR fields H$_{\text{res}}$ as a function of the IL-DMI strength for two fixed frequencies.

FMR linewidth, it does not affect the FMR frequency or field, which are the focus of the present work. Investigating the FMR linewidth [28] is beyond the scope of this paper. Therefore, the Gilbert term is not included into the model. The effective field acting on the magnetization vector is obtained as the variational derivative of the system's Hamiltonian $\mathbb{H}$ with respect to $\boldsymbol{M}_i$, taken with the negative sign $-\partial \mathbb{H}/\partial \boldsymbol{M}_i$. The dynamic part of the linearized effective field $\boldsymbol{H}_{\text{eff},i}$ comprises the field of the intralayer inhomogeneous-exchange interaction $\frac{2A_i}{\mu_0 M_{s,i}^2}\nabla^2 \boldsymbol{m}_i = \alpha_{inh,i}\nabla^2 \boldsymbol{m}_i$ ($A_i$ is the exchange stiffness, and $\alpha_{inh,i}$ is the constant of the inhomogeneous exchange interaction) and the dynamic demagnetizing (dipole) field. The dipole field has only one vector component, given by $-m_{i,x}\hat{x}$ where $\hat{x}$ is the unit vector along the x-axis, assuming that the magnetization precession is uniform in the plane of the layers, as is the case in FMR. We also account for the effective dynamic field arising from bulk in-plane uniaxial anisotropy. In this geometry, only one vector component of the linearized anisotropy field is significant: $-\frac{2K_{in}}{\mu_0 M_{s,i}^2}m_{i,y}\cdot\cos(2\varphi_{Mi})$, where $\varphi_{Mi}$ is the azimuthal vector of the anisotropy axis with respect to $\boldsymbol{M}_i$ (Fig. 1).

In addition to the fields acting within the FM layers, we also introduce effective fields at the interfaces between the magnetic layers and the NM spacers (termed "internal interfaces") by summing up all exchange contributions, including the IL-HEI field $\boldsymbol{H}_{\text{HEI}}$ and the IL-DMI field $\boldsymbol{H}_{\text{D}}$. The linearized effective dynamic Heisenberg exchange field at the interfaces is given by $\boldsymbol{H}_{\text{HEI},i} = -J_{12}(\boldsymbol{m}_j - \boldsymbol{m}_i)$, where $i \neq j$; $i,j = 1,2$. Note that the constant $J_{12}$ includes contributions from both bi-linear and bi-quadratic interlayer exchange, as the linearized effective inhomogeneous-exchange fields of both interactions exhibit the same functional dependence on $\boldsymbol{m}_1$ an $\boldsymbol{m}_2$ for our specific configuration of co-aligned static magnetization vectors in the FM layers. This is evident from our IL-BC derivation (Appendix I) and from Eq. (4) of Ref. 28. Thus, this configuration facilitates the separation of IL-DMI from IL-HEI by consolidating all IL-HEI contributions into a single $A_{12}$ constant. To derive an expression for $\boldsymbol{H}_{\text{D}}$ at the interfaces of the magnetic layers, we use the following expression for the IL-DMI energy

$$E_{\text{IL-DMI}} = -\boldsymbol{D}\cdot(\sqrt{M_{s1}^2 - m_1^2}\hat{z} + \boldsymbol{m}_1) \times (\sqrt{M_{s2}^2 - m_2^2}\hat{z} + \boldsymbol{m}_2)) \tag{2}$$

Linearizing Eq. 2 we arrive at the $\boldsymbol{H}_{\text{D}1,2}$ derived in Appendix I:

$$\boldsymbol{H}_{D1} = \frac{D_z}{\mu_0 d_1 M_{s,2}^2}(m_{2y}\hat{x} - m_{2x}\hat{y})$$

$$\boldsymbol{H}_{D2} = \frac{D_z}{\mu_0 d_1 M_{s,1}^2}(-m_{1y}\hat{x} + m_{1x}\hat{y}), \tag{3}$$

where $D_z = D\cos(\varphi_{HD})$, $D=|\boldsymbol{D}|$, and $\hat{x}$, $\hat{y}$ and $\hat{z}$ are the unit vectors along the respective axes.

The presence of IL-DMI modifies the known [24-25] IL-BC for the dynamic magnetization at the internal interfaces ($x = \pm d/2$) of MML, while the EBC at the surfaces ($x = -d_0/2 - d_1$, $x = d_0/2 + d_1$) remain the same [23-25]. The latter include torques originating solely from the intralayer inhomogeneous exchange interaction and surface anisotropy [23,25] (see also Appendix I). The linearized version of the four IL-BC is expressed as:

$$\left[\frac{A_1}{M_{s,1}}\frac{\partial m_{1y}}{\partial x} + A_{12}\frac{m_{1y}}{M_{s,1}}\right]_{x=-d_0/2} - \left[A_{12}\frac{m_{2y}}{M_{s,2}} - D_z\frac{m_{2x}}{M_{s,2}}\right]_{x=d_0/2} = 0$$

$$\left[\frac{A_1}{M_{s,1}}\frac{\partial m_{1x}}{\partial x} + (A_{12} - K_1)\frac{m_{1x}}{M_{s,1}}\right]_{x=-d_0/2} - \left[A_{12}\frac{m_{2x}}{M_{s,2}} + D_z\frac{m_{2y}}{M_{s,2}}\right]_{x=d_0/2} = 0 \tag{4}$$

$$\left[A_{12}\frac{m_{1y}}{M_{s,1}} + D_z\frac{m_{1x}}{M_{s,1}}\right]_{x=-d_0/2} + \left[\frac{A_2}{M_{s,2}}\frac{\partial m_{2y}}{\partial x} - A_{12}\frac{m_{2y}}{M_{s,2}}\right]_{x=d_0/2} = 0$$

$$\left[A_{12}\frac{m_{1x}}{M_{s,1}} - D_z\frac{m_{1y}}{M_{s,1}}\right]_{x=-d_0/2} - \left[\frac{A_2}{M_{s,2}}\frac{\partial m_{2x}}{\partial x} - (A_{12} - K_2)\frac{m_{2x}}{M_{s,2}}\right]_{x=d_0/2} = 0$$

Below, we will consider a specific set of material parameters characteristic of the Co/Pd/Co trilayer film [18,29]. However, several other material combinations, such as Co/Ag/Co or Co/Pt/Ru/Pt/CoFeB, exhibit a similar range of material parameters [9,30]. Specifically, we assume the following parameters: saturation magnetization $M_{s1} = M_{s2} = 1.393 \cdot 10^6$ A/m (17.4 kG), constants of the inhomogeneous (FM) exchange interaction $\frac{\alpha_{inh,1}}{\Delta^2} = \frac{\alpha_{inh,2}}{\Delta^2} = \frac{2 A_{1(2)}}{\Delta^2 M_{s1(2)}^2 \mu_0} =$ 73.53 with $A_1 = A_2 = 5.6 \cdot 10^{-12}$ (J/m) and the lattice constant $\Delta = 2.5 \cdot 10^{-10}$ m, IL-HEI in the range $-1.839 \leq \frac{\alpha_{12}}{\Delta} = \frac{2 A_{12}}{\Delta M_{s1(2)}^2 \mu_0} \leq 1.839$ ($-0.56 \leq A_{12} \leq 0.56$ mJ/m²), IL-DMI with $0 \leq \frac{\alpha_D}{\Delta} = \frac{2 D}{\Delta M_{s1(2)}^2 \mu_0} \leq 0.657$ ($0 \leq D \leq 0.2$ mJ/m²), perpendicular surface anisotropy $K_1 = K_2$ and respective effective fields of the anisotropy $H_{s_1} = H_{s_2} = 7010$ Oe [20]. We consider the anisotropies at the internal interfaces to be significantly stronger than those at the external surfaces [17,28] and neglect the latter.

To find the FMR fields, we introduce the general solutions of Eq. 1 $(m_{ix}(\kappa_i), m_{iy}(\kappa_i))$ [23-25], where $\kappa_i$ is the wave-vector of a standing spin wave formed across the thickness of layer $i$ [23-25]. Substituting the solutions into Eq.(1) yields two characteristic equations – one for each magnetic layer - each with four solutions for $\kappa_i$. Consequently, we have a total of eight solutions to Eq. (1). We then substitute a linear combination of these eight solutions into the EBC, resulting in a system of eight linear homogeneous equations for the coefficients of the linear combination. For nontrivial solutions of the boundary-value problem to exist, the determinant $\text{Det}[\Lambda(\omega)]$ of the matrix $\Lambda(\omega)$ of the coefficients of the system must vanish. The roots of the determinant then provide the sought values of $\kappa_i$, a total of eight – four for each magnetic layer. All solutions are functions of the FMR frequency $\omega$, generating an equation for the discrete frequencies of FMR modes at a given value of $H_0$.

The roots of the determinant can be easily found numerically in the general case. Additionally, in the special case where $A_{12}$ and $K_i$ vanish, $M_{s1} = M_{s2} = M_s$, $A_1 = A_2 = A$, a closed-form expression for the determinant can be derived. This expression represents a transcendental equation for $\kappa_i$. As a class, transcendental equations do not have exact analytical solutions; therefore, the resulting expression is not particularly useful. However, it does provide a simple and physically transparent solution for the FMR fields $H_1$, $H_2$ in the limit $D_z/A_1 = D_z/A_2 \ll 1$. The FMR fields are the values of the applied field $H_0$ necessary to tune an FMR mode's frequency to match a given frequency $\omega$. The approximate solution is derived in Appendix II. It reads:

$$H_2 - H_1 = \frac{1}{\gamma\mu_0}\frac{4\alpha_{inh}\omega_M}{d_1}\sqrt{\frac{\omega^2}{4\omega^2+\omega_M^2}}\left|\frac{D_z}{A}\right| = 2(H_2 - H_0), \qquad (5)$$

where $\omega_M = \gamma\mu_0 M_s$ and $H_0 = \frac{1}{\gamma\mu_0}\left(\sqrt{\frac{4\omega^2+\omega_M^2}{2}} - \frac{\omega_M}{2}\right)$ is the FMR field for the parent uniform mode of the single uncoupled FM layer (referred to as the "parent mode" hereafter). Equation (5) indicates that the parent mode $H_0$ must split into a doublet ($H_2$, $H_1$) in the presence of IL-DMI. In addition, we construct a numerical code similar to that of [22], treating each magnetic layer as a stack of discrete exchange-coupled microscopic "atomic" layers—a "microscopic model". This approach allows us to independently evaluate the validity of both the numerical solution of the boundary-value problem ($\text{Det}[\Lambda(\omega)] = 0$) and the approximate analytical solution given in (5).

III. **Discussion**

Figure 2a compares the numerical roots of det ($\Lambda(\omega)$) with Eq. (5), calculated using the Gauss-Newton method with a tolerance of $10^{-8}$. These traces are obtained for $\boldsymbol{D} \parallel \boldsymbol{M}_i$ and vanishing IL-HEI and anisotropies. The traces confirm that the parent mode of a single magnetic layer splits into a doublet [31]. Figure 2b demonstrates the FMR fields as a function of $D=|\boldsymbol{D}|$ [32]. As expected from Eq. (5), the mode separation $H_2$–$H_1$ increases with $D$. Furthermore, there is very good agreement between Eq. (5) and the numerical results shown in Figures 2a and 2b. Some discrepancies appear only at high values of IL-DMI, which is not unexpected since Eq. (5) was derived under the assumption of small |$\boldsymbol{D}$|. Interestingly, our numerical results show weak mode softening in the lower-frequency branch - the FMR field for $D$=0.3 J/m$^2$ and zero frequency does not vanish, remaining around 50 to 100 Oe (see the inset in Fig. 2a). However, this result should be interpreted with caution, as our model assumes perfect co-alignment of $\boldsymbol{H}_0$ with $\mathbf{M}_i$, which may not hold in real systems under small applied fields.

To understand the physical differences between the two lowest-order modes of precession, we analyse the "mode profiles," which represent the distributions of dynamic magnetization across the thickness of the structure, as shown in Appendix III. The most notable feature in these graphs is the behavior of the phases $\arg(m_{x1(2)})$, $\arg(m_{y1(2)})$ of the dynamic magnetizations. These phases exhibit jumps $\Delta\psi$ of ±π/2 at the internal interfaces (Fig. A1 in Appendix III). For the mode we will refer to as the acoustic mode (AM), the phase jumps upward ($\Delta\psi_a > 0$) when moving from the bottom layer to the top one (i.e. in the positive direction of the *x*-axis), while for the optical mode (OM), the phase jumps downward ($\Delta\psi_o < 0$). These phase jumps correspond to differences in the precession phases of the magnetization vectors in the layers. For AM, the vector $\boldsymbol{m}_1$ (bottom layer) is ahead of $\boldsymbol{m}_2$ (top layer) in the precessional trajectory, whereas for OM, $\boldsymbol{m}_1$ lags behind $\boldsymbol{m}_2$. This relative alignment of the vectors is the most important effect of IL-DMI (see Figs. A1, A2 in Appendix III).

We can now discuss our identification of these modes as acoustic and optical. The acoustic mode of oscillation in a system of two coupled resonators is conventionally defined as one in which the oscillations of the resonators are in phase ($\Delta\psi_a = 0$, "in − phase oscillations"). The optical mode (OM) is characterized by a phase difference of $\pi$ ($\Delta\psi_o = \pi$, "anti − phase oscillations"). For $A_{12} = 0$, neither mode complies with this definition. The phase difference $\Delta\psi$ is neither $-\pi/2$ or $\pi/2$. However, $\pi/2 - (-\pi/2) = \pi$, which is the same value as $\Delta\psi_a - \Delta\psi_o$ for a conventional coupled-resonantor system. Typically, optical modes (OMs) are characterized by higher oscillation energy than acoustic modes (AMs). The magnetic energy of the mode identified as "optical" in Fig. 2(a) is higher than that of the mode termed "acoustic." This is evident from the higher frequency of the optical mode compared to the acoustic mode for any applied field. Therefore, the OM is characterized by larger magnon energy. The difference in magnon energies clearly originates from IL-DMI, as all other contributions to the magnon energy are the same for both modes.

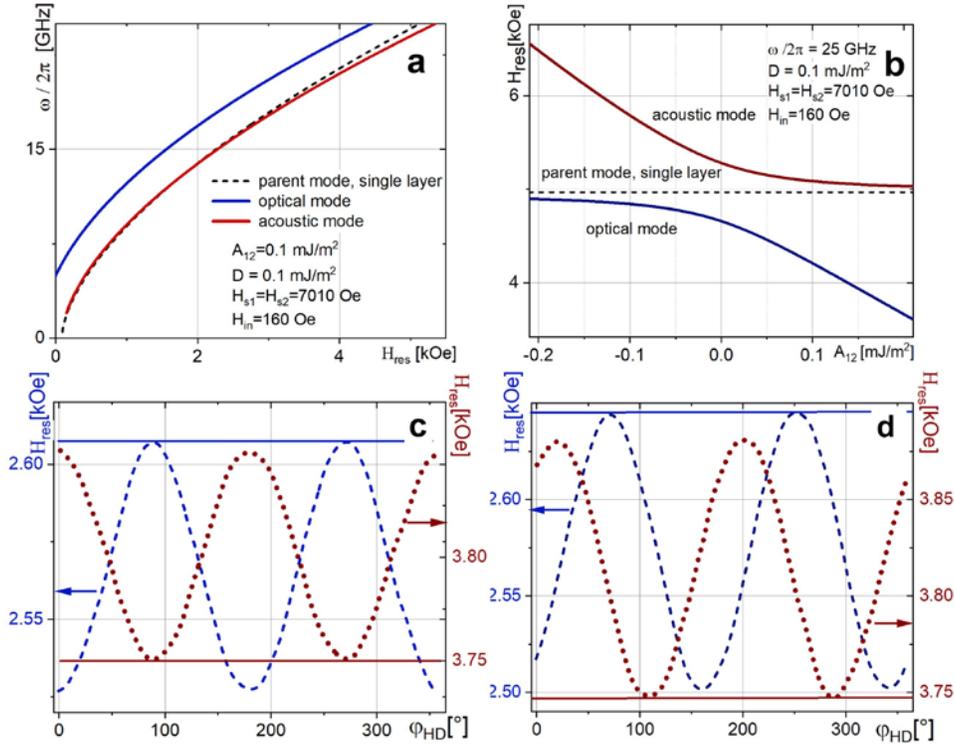

**Fig. 3**. (a) Frequency ($\omega/2\pi$) vs. resonance field ($H_{\text{res}}$) dependence of the two lowest-order resonance modes for $A_{12}$=0.1mJ/m², $D$=0.1 mJ/m², bulk in-plane anisotropy $H_{\text{in}}$=160 Oe, with the anisotropy axis parallel to the $z$-axis, and interface perpendicular anisotropy $H_{s1}$= $H_{s2}$=7010 Oe. The DMI vector $\boldsymbol{D}$ points in the positive direction of the $z$-axis. The interface anisotropy is the same for both layers and is present only at the internal layer interfaces. (b) $A_{12}$ dependence of the resonant fields for $\omega/2\pi = 25\ GHz$. The dashed lines in both panels indicate resonance frequencies or fields for the parent (quasi-)uniform mode of a single uncoupled layer. (c-d) Angular dependencies ($\varphi_{\text{HD}} = \angle(\boldsymbol{H_0}, \boldsymbol{D})$, see Fig. 1) of the FMR fields for acoustic (blue) and optical (red) modes using parameters given in (a). In (c), $H_{\text{in}}$=0. The horizontal blue and red lines correspond to FMR fields for vanishing $D$. In (d), $H_{\text{in}}$=160 Oe for both layers with the same angle between the DMI vector $\boldsymbol{D}$ and the bulk anisotropy axes $\varphi_{DK} = 35°$. The arrows in panels (c) and (d) serve to clarify which vertical axis is associated with each plot.

This understanding allows us to designate the higher frequency mode as acoustic (AM) and the lower frequency mode as optical (OM). The energy of acoustic magnons is lower because the negative angle between $\mathbf{m}_1$ and $\mathbf{m}_2$ represents the natural relative orientation of any two magnetization vectors coupled solely by IL-DMI. This orientation minimizes the energy $E_{\text{IL-DMI}}$ associated with IL-DMI while accounting for all other energy contributions. The configuration where $\mathbf{m}_1$ lags behind $\mathbf{m}_2$ is characterized by higher IL-DMI energy. Implementing this configuration requires adding extra energy to the system to force the vectors into this orientation. In an FMR experiment, this extra energy is provided by the absorbed microwave photons, necessitating a higher microwave frequency to drive the optical mode.

We now study the influence of IL-HEI and magnetic anisotropies on the FMR spectra by finding roots of $\text{Det}[\Lambda(\omega)] = 0$ numerically (Fig. 3). We consider a dynamical in-plane uniaxial anisotropy for the bulk of the material with an effective anisotropy field $\mathbf{H}_{\text{in}} = 160 \cos(2\varphi_{Mi})$ Oe [33]. It is included into both characteristic equations. Furthermore, we consider out-of-plane interfacial anisotropies with effective fields $|\mathbf{H}_{s1}| = |\mathbf{H}_{s2}| = 7010$ Oe [20] at the internal interfaces. The interfacial anisotropies at the external interfaces are assumed to vanish. These anisotropies are incorporated into the interlayer boundary conditions. Similar to the cases of non-vanishing $A_{12}$ with vanishing $D$ [24] and non-vanishing $D$ with $A_{12}=0$ (Fig. 2), we find two frequency branches located above and below the resonant field for the parent mode of a single Co layer with the same surface anisotropy (see Fig. 3a). Computing precession trajectories for the branches allows us to unambiguously identify the lower-frequency mode as the acoustic mode, as precession in the layers is nearly in-phase and becomes precisely in-phase when $D = 0$ (Fig. A2 in Appendix III). Similarly, we identify the higher-frequency mode in Fig.3a as the optical mode, where the precessional motion is close to anti-phase and becomes perfectly anti-phase for $D = 0$. The branches exhibit the same asymptote for large negative $A_{12}$ (OM) or large positive $A_{12}$ (AM). This is more clearly illustrated in Fig. 3(b), where we plot the resonant fields for the modes at a fixed frequency of 25 GHz. As shown in the panel, the asymptote corresponds to the FMR field for the parent mode [34]. These results align well with the microscopic model and those from Ref. 9.

An important question is which signatures will unambiguously identify the presence of IL-DMI in future FMR experiments. One possible answer lies in the peculiar behavior of the in-plane angle dependence of the resonance fields $H_{\text{res}}$ of the two modes denoted above as $H_1$ and $H_2$ (see Eq. 5). The IL-HEI contribution to the FMR field is independent of the angle between the applied field $\mathbf{H}_0$ and any in-plane axis. Therefore, if an angular dependence is observed experimentally, it may only originate from IL-DMI and/or the in-plane bulk anisotropy of the layers. Figure 3(c-d) illustrates the dependence of the FMR field on the angle $\varphi_{HD}$ between $\mathbf{H}_0$ and $\mathbf{D}$ [35]. One can see that the shapes of the dependencies are sine waves. A key signature of the trace in panel (c) is the "anti-phase" character of the $H_1(\varphi_{HD})$ and $H_2(\varphi_{HD})$ sine waves - the positions of the maxima for AM coincide with the positions of the minima for OM. The extrema in Fig. 3c are located at $\varphi_{HD} = n\pi/2, n = 0,1,2...$ The straight lines in panels (c-d) indicate $H_1$ and $H_2$ for $D$=0. The difference between $H_1$ and $H_2$ for $D$=0 arises from the presence of a non-vanishing $A_{12}$. Notably, the maximum FMR field for AM and the minimum field for OM coincide with their respective FMR fields for $D = 0$. This result potentially allows for the separation of $A_{12}$ from $D$.

Figure 3d displays the same traces, but this time in the presence of uniaxial easy-plane bulk anisotropy ($H_{in1} = H_{in1} = H_{in2} \approx 160$ Oe). The direction of the anisotropy axis is the same for both layers, oriented at an angle of $\varphi_{DK} = 35°$ with respect to the $D$ vector. The presence of $H_{in}$ shifts the initial phases of the angular dependencies, resulting in sine waves that are no longer in anti-phase. Importantly, for $D=0$, $H_1(\varphi_{HD})$ and $H_2(\varphi_{HD})$ are perfectly in phase (not shown). Therefore, we can interpret a non-vanishing phase difference between the two sine waves as evidence of IL-DMI in the system. This conclusion remains valid even in the absence of co-alignment between the bulk anisotropy axes of the layers (see Appendix IV).

Thus, it will be possible to infer the presence of IL-DMI in a multilayer system by conducting in-plane angle-resolved FMR measurements. If the obtained sine waves for the acoustic mode (AM) and optical mode (OM) are not in phase, it is likely that the system possesses IL-DMI. The strength of the IL-DMI can potentially be extracted by fitting the experimental dependencies to the current theory [36]. For vanishing $H_{in}$, the difference between the maximum resonance field for AM and the minimum field for OM depends solely on $A_{12}$ and can potentially be used to extract $A_{12}$ from the measurements separately from $D$. In the presence of bulk anisotropy, this is no longer the case; however, for a reasonably small anisotropy field—on the order of tens or low hundreds of Oe—the main contribution to the field difference still arises from $A_{12}$.

This method can only be employed if the optical FMR absorption peak is visible in the recorded FMR trace. Peak visibility may pose a challenge, particularly for $D=0$ and $A_{12} \neq 0$, where we might expect the amplitude of the optical peak to vanish. In a limiting case where $H_{s1}=H_{s2}=0$, $A_{12} \to \infty$ and $d_0 = 0$, the system effectively behaves like a single magnetic layer of double thickness [37]. The optical mode then reduces to the first Standing Spin Wave Mode (1st SSWM) of a single-layer structure with double thickness. Due to its perfectly anti-symmetric mode profile, the 1st SSWM is typically not observable in FMR traces (see e.g., [18]). However, the inclusion of IL-DMI alongside IL-HEI disrupts this perfect anti-symmetry, as illustrated in Fig. A2 (c) of Appendix III. This figure shows that the magnetization vector tips are not perfectly at 180° to each other, which should result in a non-vanishing amplitude of the optical absorption peak. While we do not expect the optical peak to match the amplitude of the acoustic peak, it is realistic to anticipate that for larger experimentally achievable values of $D$, the peak amplitude will exceed the noise floor of a typical FMR setup, making it reliably detectable. Furthermore, due to the high metallic conductivity of all structure layers, we can expect an even larger amplitude of the optical FMR peak when employing the Stripline Broadband FMR method [18]. Additionally, introducing asymmetries in the system—such as differences in layer interface anisotropies or thicknesses of the magnetic layers—will also enhance the visibility of the optical peak.

## ACKNOWLEDGMENTS

The authors gratefully acknowledge the University of Western Australia for the EYV's Robert and Maude Gledden Visiting Fellowship, which enabled this work. E.Y.V. also acknowledges financial support from the Deutsche Forschungsgemeinschaft (DFG) via Project No. 514141286.

solutions in Eq. (5) and two fundamental modes in the atomistic model.

[32] The larger resonance frequency for a given applied field corresponds to a smaller resonance field for a given frequency. Therefore, in Fig, 2(b), the acoustic branches are above the respective optical ones.

[33] N.G. Bebenin, A.V. Kobelev, A.P. Tankeyev, V.V. Ustinov, J. Magn. Magn. Mater. **165**, 468 (1997).

[34] When addressing the AFM IL-HE case, we assume the same FM state of static magnetization as for the FM IL-HEI scenario. This assumption implies that $H_0$ is sufficiently strong to co-align the static magnetization vectors of the AFM-coupled layers, resulting in a perfectly uniform static magnetization across the thickness of both layers, with no magnetization twists present within the layers.

[35] Recall that we assume $\boldsymbol{H}_0$ lies in the plane of the structure and is strong enough to perfectly align the static magnetization vectors of both magnetic layers along $\boldsymbol{H}_0$.

[36] The only exception to this rule occurs when the in-plane bulk anisotropy axes of the two magnetic layers are perfectly orthogonal to one another and $D=0$. In this case, the two sine waves are also in anti-phase. However, their amplitudes are very small (approximately 2 Oe for the same parameters as in Fig. 3c). Moreover, we believe the likelihood of such alignment of the axes in a real-world sample is quite low, unless it has been intentionally induced in some manner.

[37] The precise value of $d_0$ is not important within this formalism, as the theory treats $A_{12}$, $D$ and $d_0$ as independent quantities. Thus, $d_0$ serves as a purely geometric parameter that defines the co-ordinates of the interfaces without affecting the exchange conditions at those interfaces. This allows the results of the theory to remain independent of $d_0$. Consequently, $d_0$ can be assumed to take any value or set to zero to simplify formulas and numerical modelling. This simplification is applied in Eqs. (A12) below, which are presented for $d_0=0$, making them more compact without loss of generality. The strength of interlayer exchange coupling is fully captured through the values of $A_{12}$ and $D$, which specify the strength of the coupling completely.

**Appendix I. Derivation of the boundary conditions for the exchange operator that include IL-DMI**

The dynamic magnetization vector and the effective magnetic field must satisfy boundary conditions at each interface(surface) of the bilayer structure. The exchange boundary conditions applicable to the external surfaces $x=\pm(d_1 + d_0 / 2)$ are the well-established Rado-Weertmann boundary conditions for the $x$- and $y$-components of the torque at each surface (see Ref. 23-25):

$$\left[\frac{\partial m_{1y}}{\partial x}\right]_{x=-(d_1+d_0/2)} = 0 \ ;$$
$$\left[A_1 \frac{\partial m_{1x}}{\partial x} + K_1^s m_{1x}\right]_{x=-(d_1+d_0/2)} = 0, \quad (A1)$$

$$\left[\frac{\partial m_{2y}}{\partial x}\right]_{x=d_2+d_0/2} = 0,$$

$$\left[A_2 \frac{\partial m_{2x}}{\partial x} + K_2^S m_{2x}\right]_{x=d_2+d_0/2} = 0,$$

where $A_{1,2}$ is the intralayer exchange stiffness constant and $K_{1,2}^S$ is the constant of the surface perpendicular anisotropy for the respective layers.

To derive the interfacial boundary conditions for the internal interfaces $x=\pm d_0/2$, we need to include additional torques due to the interlayer exchange interaction (IL-HEI) $A_{\text{RKKY}}$, the interlayer Dzyaloshinskii-Moriya interaction (IL-DMI) $\boldsymbol{D}$, and the interlayer biquadratic exchange interaction (IL-BEI) $A_{\text{bi}}$. To this end, we employ the Hamiltonian as follows:

$$\mathcal{H} = E_{\text{IL-HEI}} + E_{\text{IL-DMI}} + E_{\text{IL-BEI}} = \\ = -A_{\text{RKKY}} \cdot (\boldsymbol{M_1} \cdot \boldsymbol{M_2}) - \boldsymbol{D} \cdot (\boldsymbol{M_1} \times \boldsymbol{M_2}) - A_{\text{bi}} \cdot (\boldsymbol{M_1} \cdot \boldsymbol{M_2})^2 \tag{A2}$$

The respective effective fields for the both layers are then given by:

$$\boldsymbol{H}_1^{eff} = -\frac{\partial \mathcal{H}}{\partial \boldsymbol{M_1}} = \boldsymbol{H}_{1RKKY}^{eff} + \boldsymbol{H}_{1D}^{eff} + \boldsymbol{H}_{1bi}^{eff}$$
$$\boldsymbol{H}_2^{eff} = -\frac{\partial \mathcal{H}}{\partial \boldsymbol{M_2}} = \boldsymbol{H}_{2RKKY}^{eff} + \boldsymbol{H}_{2D}^{eff} + \boldsymbol{H}_{2bi}^{eff}, \tag{A3}$$

where

$$\boldsymbol{H}_{1RKKY}^{eff} = \frac{A_{\text{RKKY}}}{\mu_0 d_1 M_{s,1} M_{s,2}} (M_{2x}, M_{2y}, M_{2z})$$
$$\boldsymbol{H}_{2RKKY}^{eff} = \frac{A_{\text{RKKY}}}{\mu_0 d_1 M_{s,1} M_{s,2}} (M_{1x}, M_{1y}, M_{1z})$$
$$\boldsymbol{H}_{1D}^{eff} = \frac{1}{d_1 \mu_0 M_{s,1} M_{s,2}} (D_z M_{2y} - D_y M_{2z}, -D_z M_{2x} + D_x M_{2z}, D_y M_{2x} - D_x M_{2y}) \tag{A4}$$
$$\boldsymbol{H}_{2D}^{eff} = \frac{1}{d_1 \mu_0 M_{s,1} M_{s,2}} (-D_z M_{1y} + D_y M_{1z}, D_z M_{1x} - D_x M_{1z}, -D_y M_{1x} + D_x M_{1y})$$
$$\boldsymbol{H}_{1bi}^{eff} = \frac{2 A_{\text{bi}}}{\mu_0 d_1 M_{s,1} M_{s,2}} ((M_{2x}(M_{1x}M_{2x} + M_{1y}M_{2y}), M_{2y}(M_{1x}M_{2x} + M_{1y}M_{2y}), M_{2z}(M_{1x}M_{2x} + M_{1y}M_{2y}))$$
$$\boldsymbol{H}_{2bi}^{eff} = \frac{2 A_{\text{bi}}}{\mu_0 d_1 M_{s,1} M_{s,2}} ((M_{1x}(M_{1x}M_{2x} + M_{1y}M_{2y}), M_{1y}(M_{1x}M_{2x} + M_{1y}M_{2y}), M_{1z}(M_{1x}M_{2x} + M_{1y}M_{2y})),$$

$M_{1x}$, $M_{2x}$, $M_{1y}$, $M_{2y}$, $M_{1z}$, $M_{2z}$, $D_x$, $D_y$ and $D_z$ are the projections of the respective vectors on the axes $x$, $y$ and $z$, and $M_{s,1} M_{s,2}$ are the saturation magnetizations for the two layers.

Recall that we consider the case where the applied field is strong enough to co-align the static magnetization vectors of both layers with the applied field, that is to orient them perfectly along the z-axis. In this case, the *x*- and *y*-components of the magnetization vector are dynamic. Therefore, we replace them with dynamic magnetization components $m_{1,2x}$ and $m_{1,2y}$ : $\boldsymbol{M_{1,2}} = (m_{1,2x}, m_{1,2y}, M_{1,2z})$, where $M_{1,2z} = \sqrt{M_0^2 - m_{1,2x}^2 - m_{1,2y}^2}$. The dynamic components are small compared to $M_{1,2z}$: $|m_{1,2x}|, |m_{1,2y}| \ll M_{1,2z}$. We can then use the derived formulas to express the magnetic torques acting on the two layers:

$$\boldsymbol{T}_1^{eff} = \boldsymbol{M_1} \times \boldsymbol{H}_{1RKKY}^{eff} + \boldsymbol{M_1} \times \boldsymbol{H}_{1D}^{eff} + \boldsymbol{M_1} \times \boldsymbol{H}_{1bi}^{eff}$$

$$\boldsymbol{T}_2^{eff} = \boldsymbol{M}_2 \times \boldsymbol{H}_{2RKKY}^{eff} + \boldsymbol{M}_2 \times \boldsymbol{H}_{2D}^{eff} + \boldsymbol{M}_2 \times \boldsymbol{H}_{2bi}^{eff} \tag{A5}$$

Once the torques are evaluated, we linearize all expressions by neglecting terms of second order in smallness. This process results in the following expressions for the torques:

$$\begin{aligned}
\boldsymbol{T}_1^{RKKY} &= A_{\text{RKKY}}\left(\frac{m_{1y}}{M_{s,1}} - \frac{m_{2y}}{M_{s,2}}, -\frac{m_{1x}}{M_{s,1}} + \frac{m_{2x}}{M_{s,2}}, 0\right) \\
\boldsymbol{T}_2^{RKKY} &= A_{\text{RKKY}}\left(-\frac{m_{1y}}{M_{s,1}} + \frac{m_{2y}}{M_{s,2}}, \frac{m_{1x}}{M_{s,1}} - \frac{m_{2x}}{M_{s,2}}, 0\right) \\
\boldsymbol{T}_1^D &= D_z\left(\frac{m_{2x}}{M_{s,2}}, \frac{m_{2y}}{M_{s,2}}, 0\right) \\
\boldsymbol{T}_2^D &= D_z\left(-\frac{m_{1x}}{M_{s,1}}, -\frac{m_{1y}}{M_{s,1}}, 0\right) \\
\boldsymbol{T}_1^{bi} &= 2A_{\text{bi}}\left(\frac{m_{1y}}{M_{s,1}} - \frac{m_{2y}}{M_{s,2}}, -\frac{m_{1x}}{M_{s,1}} + \frac{m_{2x}}{M_{s,2}}, 0\right) \\
\boldsymbol{T}_2^{bi} &= 2A_{\text{bi}}\left(-\frac{m_{1y}}{M_{s,1}} + \frac{m_{2y}}{M_{s,2}}, \frac{m_{1x}}{M_{s,1}} - \frac{m_{2x}}{M_{s,2}}, 0\right)
\end{aligned} \tag{A6}$$

In equilibrium, all torques must vanish. This leads to the desired boundary conditions. In addition to the interlayer-exchange torques, the boundary conditions must also incorporate the usual "Rado-Weertmann torques". These torques are the same as those that yield Eq.(A1). Notably, in our geometry, the expressions for the torques arising from $A_{\text{RKKY}}$ and $A_{\text{bi}}$ share the same form, differing only by the coeffient in front of the expression. The torques combine, resulting in a single term with a coefficient $A_{12} = A_{\text{RKKY}} + 2A_{\text{bi}}$ in front of it. Consequently, in this particular geometry, the bi-linear and Heisenberg exchange interactions are indistinguishable.
The derived boundary conditions are as follows:

$$\left[\frac{A_1}{M_{s,1}}\frac{\partial m_{1y}}{\partial x} + A_{12}\frac{m_{1y}}{M_{s,1}}\right]_{x=-d_0/2} - \left[A_{12}\frac{m_{2y}}{M_{s,2}} - D_z\frac{m_{2x}}{M_{s,2}}\right]_{x=d_0/2} = 0$$

$$\left[\frac{A_1}{M_{s,1}}\frac{\partial m_{1x}}{\partial x} + (A_{12} - K_1)\frac{m_{1x}}{M_{s,1}}\right]_{x=-d_0/2} - \left[A_{12}\frac{m_{2x}}{M_{s,2}} + D_z\frac{m_{2y}}{M_{s,2}}\right]_{x=d_0/2} = 0 \tag{A7}$$

$$\left[A_{12}\frac{m_{1y}}{M_{s,1}} + D_z\frac{m_{1x}}{M_{s,1}}\right]_{x=-d_0/2} + \left[\frac{A_2}{M_{s,2}}\frac{\partial m_{2y}}{\partial x} - A_{12}\frac{m_{2y}}{M_{s,2}}\right]_{x=d_0/2} = 0$$

$$\left[A_{12}\frac{m_{1x}}{M_{s,1}} - D_z\frac{m_{1y}}{M_{s,1}}\right]_{x=-d_0/2} - \left[\frac{A_2}{M_{s,2}}\frac{\partial m_{2x}}{\partial x} - (A_{12} - K_2)\frac{m_{2x}}{M_{s,2}}\right]_{x=d_0/2} = 0$$

**Appendix II: Derivation of the expressions for the FMR fields**

The presence of the exchange operator $\alpha_{inh,i}\nabla^2\boldsymbol{m}_i$ in the effective field that appears in Eq.(1) transforms Eq.(1) into a differential equation in the variable $\boldsymbol{m}_i$. Its solution can be expressed in the following form:

$$\boldsymbol{m}_i(x) \sim exp(\mathbb{i}\, Q\, x), \tag{A8}$$

where $\mathbb{i}$ is the imaginary unit.

Substituting the solution into Eq.(1) results in a characteristic equation

$$Q^2 \alpha^2 \omega_M^2 + 2 Q \omega_H \omega_M + Q \alpha \omega_M^2 - \omega^2 + \omega_H^2 + \omega_H \omega_M = 0 \qquad (A9)$$

where $\omega$ is the magnetization precession frequency, $\omega_H = \gamma \mu_0 H$, $\omega_M = \gamma \mu_0 M_{1(2)}$ and $\gamma$ is the gyromagnetic ratio. Solving this equation with respect to $Q$ yields four routs $\pm Q_1$ and $\pm Q_2$:

$$Q_1 = -\frac{2\omega_H + \omega_M - \sqrt{4\omega^2 + \omega_M^2}}{2\alpha_{inh}\omega_M}, \quad Q_2 = -\frac{2\omega_H + \omega_M + \sqrt{4\omega^2 + \omega_M^2}}{2\alpha_{inh}\omega_M}, \qquad (A10)$$

where $\alpha_{inh} \to \frac{A_{1(2)} 8\pi}{M_{1(2),z}^2} A_{1(2)}$ is the inhomogeneous-exchange constant for the material of the bulk of the corresponding layer. The total solution to Eq.(1) then reads:

$$\begin{aligned}
m_{1x} &= C_1 e^{i(\sqrt{Q_1})x} + C_2 e^{-i(\sqrt{Q_1})x} + C_3 e^{i(\sqrt{Q_2})x} + C_4 e^{-i(\sqrt{Q_2})x} \\
m_{2x} &= C_5 e^{i(\sqrt{Q_1})x} + C_6 e^{-i(\sqrt{Q_1})x} + C_7 e^{i(\sqrt{Q_2})x} + C_8 e^{-i(\sqrt{Q_2})x} \\
m_{1y} &= 2i\omega a_1 C_1 e^{i(\sqrt{Q_1})x} + 2i\omega a_1 C_2 e^{-i(\sqrt{Q_1})x} + 2i\omega a_2 C_3 e^{i(\sqrt{Q_2})x} + 2i\omega a_2 C_4 e^{-i(\sqrt{Q_2})x} \\
m_{2y} &= 2i\omega a_1 C_5 e^{i(\sqrt{Q_1})x} + 2i\omega a_1 C_6 e^{-i(\sqrt{Q_1})x} + 2i\omega a_2 C_7 e^{i(\sqrt{Q_2})x} + 2i\omega a_2 C_8 e^{-i(\sqrt{Q_2})x},
\end{aligned} \qquad (A11)$$

where $a_1 = \frac{\omega}{\omega_M - \sqrt{4\omega^2 + \omega_M^2}}$ and $a_2 = \frac{\omega}{\omega_M + \sqrt{4\omega^2 + \omega_M^2}}$. The latter two quantities are obtained from Eq. (1) with Eq.(A8).

We now substitute Eqs. (A11) into the IL-BC - Eqs. (A7) (Eqs. (4) from the main text). After some algebra, we obtain the following vector-matrix equation:

$$\begin{pmatrix}
-a_1 d_1 e^{-id_1\sqrt{Q_1}}\sqrt{Q_1} & a_1 d_1 e^{id_1\sqrt{Q_1}}\sqrt{Q_1} & -a_2 d_1 e^{-id_1\sqrt{Q_2}}\sqrt{Q_2} & a_2 d_1 e^{id_1\sqrt{Q_2}}\sqrt{Q_2} & 0 & 0 & 0 & 0 \\
d_1 e^{-id_1\sqrt{Q_1}}\sqrt{Q_1} & -d_1 e^{id_1\sqrt{Q_1}}\sqrt{Q_1} & d_1 e^{-id_1\sqrt{Q_2}}\sqrt{Q_2} & -d_1 e^{id_1\sqrt{Q_2}}\sqrt{Q_2} & 0 & 0 & 0 & 0 \\
-2a_1 d_1\sqrt{Q_1} & 2a_1 d_1\sqrt{Q_1} & -2a_2 d_1\sqrt{Q_2} & 2a_2 d_1\sqrt{Q_2} & d_1\,DA1 & d_1\,DA1 & d_1\,DA1 & d_1\,DA1 \\
d_1\sqrt{Q_1} & -d_1\sqrt{Q_1} & d_1\sqrt{Q_2} & -d_1\sqrt{Q_2} & -2a_1 d_1\,DA1 & -2a_1 d_1\,DA1 & -2a_2 d_1\,DA1 & -2a_2 d_1\,DA1 \\
d_1\,DA2 & d_1\,DA2 & d_1\,DA2 & d_1\,DA2 & -2a_1 d_1\sqrt{Q_1} & 2a_1 d_1\sqrt{Q_1} & -2a_2 d_1\sqrt{Q_2} & 2a_2 d_1\sqrt{Q_2} \\
-2a_1 d_1\,DA2 & -2a_1 d_1\,DA2 & -2a_2 d_1\,DA2 & -2a_2 d_1\,DA2 & d_1\sqrt{Q_1} & -d_1\sqrt{Q_1} & d_1\sqrt{Q_2} & -d_1\sqrt{Q_2} \\
0 & 0 & 0 & 0 & -a_1 d_1 e^{id_1\sqrt{Q_1}}\sqrt{Q_1} & a_1 d_1 e^{-id_1\sqrt{Q_1}}\sqrt{Q_1} & -a_2 d_1 e^{id_1\sqrt{Q_2}}\sqrt{Q_2} & a_2 d_1 e^{-id_1\sqrt{Q_2}}\sqrt{Q_2} \\
0 & 0 & 0 & 0 & d_1 e^{id_1\sqrt{Q_1}}\sqrt{Q_1} & -d_1 e^{-id_1\sqrt{Q_1}}\sqrt{Q_1} & d_1 e^{id_1\sqrt{Q_2}}\sqrt{Q_2} & -d_1 e^{-id_1\sqrt{Q_2}}\sqrt{Q_2}
\end{pmatrix} \begin{pmatrix} C_1 \\ C_2 \\ C_3 \\ C_4 \\ C_5 \\ C_6 \\ C_7 \\ C_8 \end{pmatrix} = 0$$

(A12)

where $DA1 = \frac{D_z M_{1,z}}{A_1 M_{2,z}}$, $DA2 = \frac{D_z M_{2,z}}{A_2 M_{1,z}}$, and we intentionnaly mulitplied every line of the system by $d_1$ to compensate for very large values of $Q_2$. The latter is important for the numerical evaluation of the determinant of the equation system.

The determinant of the 8×8 matrix must vanish to allow non-trivial solutions to the equation. This condition represents an implicit equation for the FMR frequencies or fields. For $A_{12} \to 0$, the equation can be reduced to a compact transcendental equation. For a given $\omega$, the equation roots correspond to the FMR fields $\omega_H$. In a limiting case, where $D_z/A_1 = D_z/A_2 \ll 1$, the transcendental equation admits two approximate solutions - $H_1 = \omega_{H_1}/\gamma$, and $H_2 = \omega_{H_2}/\gamma$. They correspond to the resonance fields for the optical and the acoustic modes repectively. The solution is shown in the main text as Eq. (5):

$$H_2 - H_1 = \frac{1}{\gamma\mu_0}\frac{4\alpha_{inh}\omega_M}{d_1}\sqrt{\frac{\omega^2}{4\omega^2+\omega_M^2}}\left|\frac{D_z}{A}\right| = 2(H_2 - H_0),$$

where $H_0 = \frac{1}{\gamma\mu_0}\left(\sqrt{\frac{4\omega^2+\omega_M^2}{2}} - \frac{\omega_M}{2}\right)$ is the FMR field for the parent uniform mode of the single uncoupled FM layer.

For larger values of $D$ and for $A_{12} \neq 0$, the implicit equation for the FMR fields must be solved numerically.

## Appendix III. Mode profiles

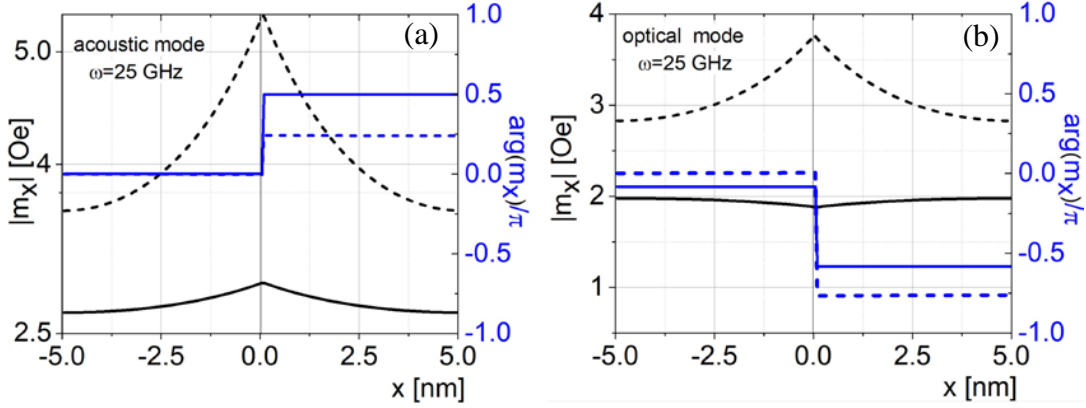

**Fig. A1** Profiles of the fundamental FMR modes for two sets of energy parameters $A_{12}$ and $D$ across the bilayer. Solid curves correspond to $D = 0.1$ mJ/m² and $A_{12}=0$, while dashed curves correspond to $D = A_{12} = 0.1$ mJ/m². Panels (a) and (b) display the amplitude (left axes, black lines) and phase (right axes, blue lines) profiles of magnetization. In both cases, the bulk in-plane anisotropy is assumed to be the same for both layers with an anisotropy field $H_{in}=160$ Oe and the anisotropy axis parallel to the $z$-axis. For clarity, interface and surface perpendicular anisotropies are neglected. To enhance clarity, the spacer is collapsed into a single point at $x=0$ (as explained in the text).

Fig. A1 shows the mode profiles for the optical mode and the acoustic mode. The profiles represent the distributions of the amplitude of the dynamic magnetization across the thickness of the structure. Although his calculation assumes a non-vanishing $d_0$ [37], the non-magnetic spacer is omitted in these plots for clarity, effectively collapsing it into an interface at $x=0$. The most striking feature in these graphs is the behavior of the phases $\varphi(m_x)$, $\varphi(m_y)$ of the dynamic components of the magntization vector. When only interlayer Dzyaloshinskii-Moriya interaction (IL-DMI) is present, with interlayer Heisenberg exchange (IL-HE) vanishing (panel (a)), the phases exhibit jumps of $\Delta\varphi = \pm\pi/2$ at the interface between the two magnetic layers. In the presence of non-vanishing $A_{12}$, (panel (b)), while phase jumps still occur, their magnitudes change significantly.

In order to understand the physical significance of the phase jumps, it is instructive to convert the $m_x$ and $m_y$ profiles into precession trajectories. A trajectory represents the geometric figure traced by the tip of the magnetization vector as it undergoes precessional motion. The trajectories for the acoustic and optical modes derived from panels (a) and (b) are shown in Fig. A2. In these graphs, the sense of precession is clockwise.

One can observe that the precession is highly elliptical, with the dynamic magnetization $m_x$ perpendicular to the bilayer structure plane being about two times smaller than the in-plane component $m_y$. This behavior is expected for thin layers of Co, attributed to cobalt's large saturation magnetization and the shape anisotropy associated with the thin-film geometry. The dashed lines represent the trajectories of the tip of **m** at the internal interface of the first (lower) magnetic layer ($x=-d_0/2$), while the solid lines correspond to the internal interface of the second (upper) magnetic layer ($x=d_0/2$).

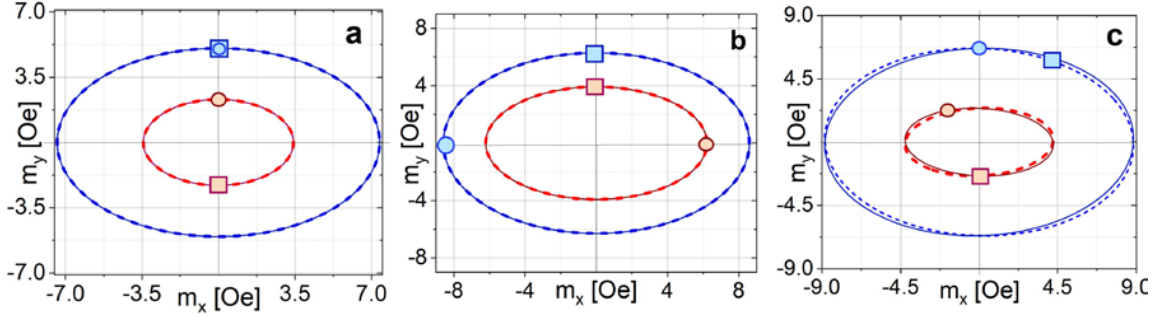

**Fig. A2** Precession trajectories for the acoustic (blue lines) and the optical (red lines) modes. Dashed lines represent the trajectories at the internal interface of Layer 1 (the bottom layer), while solid lines depict the trajectories at the internal interface of Layer 2 (the top layer). Circles and rectangles indicate the positions of the tip of the magnetization vector at a specific moment in time. Circles correspond to the lower-layer trajectories, while rectangles correspond to the upper-layer trajectories. In all cases, the precession freqeuncy is set to $\omega = 25$ GHz, the surface anisotropy field is $H_K = 7010$ Oe, and the bulk anisotropy field is $H_{in} = 160$ Oe, with the anisotropy axis parallel to the $z$-axis. In (a) $A_{12} = 0.2$ mJ/m$^2$, $D = 0$; in (b) $A_{12} = 0$, $D = 0.2$ mJ/m$^2$ ; in (c) $A_{12} = 0.4$ mJ/m$^2$, $D = 0.2$ mJ/m$^2$. It's important to emphasize that the amplitude of precession is arbitrary and carries no physical significance (unless the system is driven by an external signal). This observation allows us to fit trajectories for both AM and OM into the same graph by intentionally adjusting the radii of the trajectories for the optical mode to be smaller than those for the acoustic mode.

The blue color corresponds to the acoustic mode of precession, while the red color denotes the optical mode. Notably, the precession trajectories are identical for both layers in panels (a) and (b). This is consistent with the equal values of the magnetization vector components at the internal interfaces, $|m_x(d_0/2)|=|m_x(-d_0/2)|$ and $|m_y(d_0/2)|=|m_y(-d_0/2)|$ as seen in Fig. A1. (Recall that that the spacer is not shown in Fig. A1, so $d_0/2 = -d_0/2 = 0$ in Fig. A1.)

The circles and squares indicate the instantaneous positions of the respective vector tips at the same moment in time. One can see that **m**$_1$ is ahead of **m**$_2$ for the AM when $A_{12} = 0$ and $D = 0.2$ mJ/m$^2$ >0 (panel (b)). Conversely, **m**$_1$ lags behind **m**$_2$ for the OM (the same panel). In contrast, in panel (a), where IL-DMI is absent, the vectors are perfectly co-aligned for any moment in time for the acoustic mode (AM), indicating in-phase precession, and perfectly anti-aligned for the optical mode, indicating anti-phase precession. The lack of collinearity of the dynamic magnetization vectors is the key effect of interlayer Dzyaloshinskii-Moriya interaction (IL-DMI). It underlies all other aspects of magnetization dynamics in IL-DMI coupled multilayer magnetic structures.

Let us now discuss how this picture changes when both the interlayer Heisenberg exchange interaction and the interlayer DMI are present. Panel (c) of Fig. A2 illustrates the precession trajectories for the case where $A_{12}$>0, $D$>0. Here, we can see that the addition of ferromagnetic IL-HEI causes the ellipses to tilt relative to the multilayer structure plane. This introduces a symmetry breaking due to the combined effects of the two interactions.

Additionally, we observe that the optical mode is now characterized by a nearly anti-phase precession. This aligns with the perfectly anti-phase precession expected for the optical mode when $D = 0$, $A_{12} > 0$ (as seen in panel (a)). Importantly, its frequency is larger than that of the mode exhibiting nearly in-phase precession in panel (c). For the latter mode, the phase difference becomes precisely zero when $D=0$ (panel (a)). Both the lower frequency and the nearly zero phase difference identify this mode as acoustic. This supports our previous identification of the modes for $A_{12}=0$ (panel (b)) based on their resonance fields.

Changing sign of $D$ to negative (for $A_{12}=0$) reverses the sign of the angle between $\mathbf{m}_1$ and $\mathbf{m}_2$. Now $\mathbf{m}_2$ leads $\mathbf{m}_1$ for the acoustic mode and lags behind $\mathbf{m}_1$ for the optical one. Introducing a positive $A_{12}$ brings the angle close to 180 degree for the optical mode. However, in this case, the elipses are tilted in the opposite senses compared to panel (c) of Fig. 2.

**Appendix IV. Additional details of the angle dependence of the resonance field**

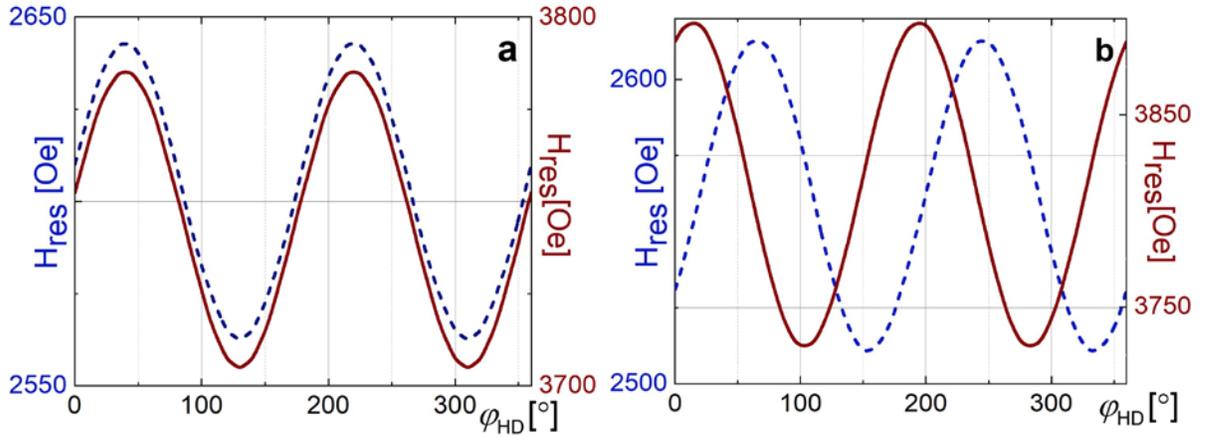

**Fig. A3** Angular dependencies of the FMR fields for acoustic (blue dashed lines) and optical (brown solid lines) modes. (a) corresponds to $D=0$, $A_{12}=0.2$ mJ/m$^2$, $H_K=0$ and $H_{in}=160$ Oe. The angle between the $H_{in}$ of the bottom layer with respect to $\mathbf{D}$ is $\varphi_{DK1} = 35°$, while that for the top layer $\varphi_{DK2} = 65°$. In (b) $D=0.2$ mJ/m$^2$. All other energy parameters are the same as in (a).

In Fig. A3 we investigate the effect of misalignment between the axes of bulk uniaxial anisotropy in the magnetic layers. The first case considered is $D=0$, $A_{12}=0.2$ mJ/m$^2 > 0$, with a 30-degree angle between the anisotropy axes of layers 1 and 2 (Panel (a)). The anisotropy axis of the first layer is oriented at 35 degrees to $\mathbf{D}$, while the second-layer's anisotropy axis is at 65 degrees to the same vector. Our modeling results in in-phase sine waves for the angular dependences of the resonance fields of both the acoustic and the optical mode. These sine waves exhibit the same amplitude for both modes. The first maximum of the sine waves corresponds to $\varphi_{HD} = 42°$. This situation is almost indistinguishable from the case where both anisotropy axes are co-aligned at 42 degrees to $\mathbf{D}$, as demonstrated by our numerical modelling. Introducing a non-vanishing $\mathbf{D}$ results in angular dependences (Panel (b)) similar to those shown in Fig. 3(d). The dependencies are characterized by a non-vanishing phase difference for the sine waves. While the overall shapes of the traces remain comparable to the case with co-aligned axes at 42 degrees to $\mathbf{D}$, the amplitude of the sine wave for the optical mode is now much closer to that of the acoustic mode compared to the 42-degree alignment case.

This behavior is consistent across all angles of misalignment—from 0 to 90 degrees. Notably, for perfectly perpendicular anisotropy axes of the two layers (i.e., at 90 degrees to each other) with $D$=0, the resulting traces are very similar to those presented in Fig. 3(c). However, the amplitude of the sine waves in this case is significantly smaller (only 2 Oe). This notably reduced amplitude is in agreement with our expectation that the presence of bulk uniaxial anisotropies in the layers should not noticeably affect the FMR fields in this very specific case.

One more special case is also worth discussing: the perfect co-alignment of the anisotropy axes of both layers with $D$ ($\varphi_{DK1} = \varphi_{DK2} = 0$). Our calculations yield anti-phase angular dependencies that are qualitatively very similar to those shown in Fig. 3(c). The anti-phase behaviour arises from the presence of a non-vanishing IL-DMI. While, as before, one can infer the presence of IL-DMI from the opposite-phase oscillations, a visual inspection of the plot alone is insufficient to ascertain whether $H_{in} \neq 0$ in this case. However, there is a quantitative difference from the previous case that may potentially be exploited to determine whether a non-vanishing $H_{in}$ is present in the magnetic layers. In the limiting case of negligible IL-DMI energy compared to the bulk anisotropy energy, the dynamic magnetization vectors for the two layers must co-align, resulting in two sine waves that perfectly overlap each other. This suggests that the sine-wave amplitude carries information about the presence of the two concurrent interactions. Fitting the perfect anti-phase dependence with the numerical model may provide a way to separate the bulk anisotropy from IL-DMI in this special case of $\varphi_{DK1} = \varphi_{DK2} = 0$.